\documentclass{nature}
\usepackage{graphicx, color}
\usepackage{amsfonts,amssymb,amsmath}
\usepackage{siunitx}

\title{Proximity-induced superconductivity in (Bi$_{1-x}$Sb$_x$)$_2$Te$_3$ topological-insulator nanowires}

\author{Mengmeng Bai$^{1,4}$, Xian-Kui Wei$^{2,4}$, Junya Feng$^{1,4}$, Martina Luysberg$^{2}$, Andrea Bliesener$^{1}$,\\
 Gertjan Lippertz$^{1,3}$, Anjana Uday$^{1}$, Alexey A. Taskin$^1$, Joachim Mayer$^{2}$ \& Yoichi Ando$^{1,*}$}

\makeatletter
\let\saved@includegraphics\includegraphics
\AtBeginDocument{\let\includegraphics\saved@includegraphics}
\renewenvironment*{figure}{\@float{figure}}{\end@float}
\makeatother

\begin{document}

\maketitle

\begin{affiliations}
 \item Physics Institute II, University of Cologne, Z\"ulpicher Str. 77, 50937 K\"oln, Germany.
 \item Ernst Ruska-Centre for Microscopy and Spectroscopy with Electrons, Forschungszentrum J\"ulich GmbH, 52425 J\"ulich, Germany.
 \item KU Leuven, Quantum Solid State Physics, Celestijnenlaan 200 D, 3001 Leuven, Belgium.
 \item These authors contributed equally. \\
 $^*${\rm e-mail: ando@ph2.uni-koeln.de}
\end{affiliations}

\begin{abstract}
\section*{Abstract}

When a topological insulator is made into a nanowire, the interplay between topology and size quantization gives rise to peculiar one-dimensional states whose energy dispersion can be manipulated by external fields. In the presence of proximity-induced superconductivity, these 1D states offer a tunable platform for Majorana zero modes. While the existence of such peculiar 1D states has been experimentally confirmed, the realization of robust proximity-induced superconductivity in topological-insulator nanowires remains a challenge. Here, we report the realization of superconducting topological-insulator nanowires based on (Bi$_{1-x}$Sb$_x$)$_2$Te$_3$ (BST) thin films. When two rectangular pads of palladium are deposited on a BST thin film with a separation of 100-200 nm, the BST beneath the pads is converted into a superconductor, leaving a nanowire of BST in-between. We found that the interface is epitaxial and has a high electronic transparency, leading to a robust superconductivity induced in the BST nanowire. Due to its suitable geometry for gate-tuning, this platform is promising for future studies of Majorana zero modes. 

\end{abstract}

%
%
\section*{Introduction}

The decisive characteristic of a topological insulator (TI) is the existence of gapless surface states whose gapless nature is protected by time reversal symmetry. However, in a TI nanowire, the size quantization effect turns the topological surface states into peculiar one-dimensional (1D) states that are gapped due to the formation of subbands\cite{Zhang2010, Bardarson2010}. Interestingly, their energy dispersion can be manipulated by both magnetic and electric fields, and these tunable 1D states have been proposed to be a promising platform to host Majorana zero modes (MZMs) in the presence of proximity-induced superconductivity\cite{Cook2011, Legg2021}. Therefore, realization of robust superconductivity in TI nanowires using the superconducting proximity effect is important for the future Majorana research. So far, the existence of the 1D subbands in TI nanowires has been probed by various means\cite{Cho2015, Jauregui2016, Dufouleur2017, Ziegler2018, Muenning2021, Rosenbach2021b, Legg2021b}, but it has been difficult to induce robust superconductivity in TI nanowires\cite{Rosenbach2021}, because a suitable method to realize an epitaxial interface between a TI nanowire and a superconductor has been lacking. 

In terms of the realization of an epitaxial interface, it was previously reported\cite{Bai2020} that when Pd is sputter-deposited on a (Bi$_{1-x}$Sb$_x$)$_2$Te$_3$ (BST) thin film, the diffusion of Pd into BST at room temperature results in a self-formation of PdTe$_2$ superconductor (SC) in the top several quintuple-layers (QLs) of BST, leaving an epitaxial horizontal interface. Within the self-formed PdTe$_2$ layer, Bi and Sb atoms remain as substitutional or intercalated impurities. This self-epitaxy was demonstrated to be useful for the fabrication of TI-based superconducting nanodevices involving planar Josephson junctions\cite{Bai2020}. In such devices, the BST remains a continuous thin film and the superconducting PdTe$_2$ proximitizes the surface states from above. In the present work, we found that depending on the Pd deposition method, Pd can diffuse deeper into the BST film and the conversion of BST into a PdTe$_2$-based superconductor can take place in the whole film. Specifically, this full conversion occurs when Pd is thermally-deposited onto BST films with the thickness up to $\sim$30 nm. This finding opened the possibility to create a TI nanowire sandwiched by SCs, by converting most of the BST film into the SC and leaving only a $\sim$100-nm-wide BST channel to remain as the pristine TI. This way, one may induce robust superconductivity in the BST naowire proximitized by the PdTe$_2$ superconductor on the side through an epitaxial interface.

Furthermore, this structure allows for gate-tuning of the chemical potential in the BST nanowire from both top and bottom sides (called dual-gating), which offers a great advantage to be able to independently control the overall carrier density and the internal electric field\cite{Liu2014}. This is useful, because the realization of Majorana zero modes (MZMs) in a TI nanowire requires to lift the degeneracy of the quantum-confined subbands\cite{Cook2011, Legg2021}, which can be achieved either by breaking time-reversal symmetry (by threading a magnetic flux along the nanowire)\cite{Cook2011} or by breaking inversion symmetry (by creating an internal electric field)\cite{Legg2021}. 

In this paper, we show that the devices consisting of BST nanowire sandwiched by PdTe$_2$ superconductor indeed present robust proximity-induced superconductivity, evidenced by signatures of multiple Andreev reflections. Furthermore, in our experiment of the ac Josephson effect, we observed that the first Shapiro step is systematically missing at low rf frequency, low rf power, and at low temperature, which points to 4$\pi$-periodic current-phase relation and gives possible evidence for Majorana bound states. This result demonstrates that the superconducting TI nanowire realized here is a promising platform for future studies of MZMs.

\section*{Results}

\subsection{Nanowire formation.} 
To confirm whether TI nanowires can indeed be created by Pd diffusion, we prepared BST/Pd bilayer samples specially designed  for scanning tunneling electron microscope (STEM) studies: a $\sim$30-nm-thick BST film was epitaxially grown on an InP (111)A substrate and $\sim$20-nm-thick rectangular Pd pads were thermally-deposited on BST with distances of $\sim$200 nm between them. 
Figure 1a shows a high-angle annular-dark-field (HAADF) STEM image of the cross-section of such a sample; here, the two Pd pads are separated by 231.7 nm. The corresponding elemental maps of Pd, Bi, Sb and Te detected by energy-dispersive X-ray (EDX) spectroscopy (Figs. 1b-1e) show the Pd not only penetrates fully into the BST film beneath the pad, but also propagates outward by $\sim$90 nm from each edge (Fig. 1b); note that in this sample, the outward diffusion was along the  $\langle110\rangle$ direction (see Fig. 1e inset). As a result, a BST nanowire with the width of $\sim$40 nm is left in-between the two Pd-containing regions. One can also see the swelling of Pd-absorbed BST (hereafter called Pd-BST) along the out-of-plane direction compared to the pristine BST region. 
We note that this STEM observation was made about 3 months after the Pd deposition, so that the in-plane Pd-diffusion length of $\sim$90 nm is probably longer than that in devices that were measured soon after the Pd deposition. For additional STEM data for various Pd-pad separations, see Supplementary Note 1 and Supplementary Fig. 1.

After confirming the creation of a TI nanowire, the next important question is the morphology of the system and identification of the converted phase. First, atomic-resolution examination of the TI nanowire region (Fig. 1f) confirms high-quality epitaxial growth of BST on the InP (111) surface, with QLs remain intact. Near the interface between BST and Pd-BST (Fig.~1g and Supplementary Fig. 2), one can see that the existence of Pd completely transforms QLs into a triple-layer (TL) structure characteristic of the PdTe$_2$-like phase (Fig. 1i)\cite{Bai2020}. Interestingly, the forefront of Pd-BST forms a V-shaped ``epitaxial'' interface to make a compromise between the different thicknesses of the QL and the TL. In the Pd-BST region closer to the Pd pad (Fig. 1h), we found that the TL structure changes to a zigzag pattern (Fig. 1k). Our image-simulation-based atomic structure analysis suggests that the TL- and zigzag-structured Pd-BST correspond to PdTe$_2$-like and PdTe-like structural phases (Fig. 1j), respectively\cite{finlayson1986}. Depending on the concentration of Pd at different locations, the chemical formula may vary between Pd(Bi,Sb,Te)$_2$ and Pd(Bi,Sb,Te), possibly with partial occupation of the Pd sites by Sb as well. Note that both PdTe$_2$ and PdTe are SCs, with reported $T_c$ of 2.0 and 4.5 K, respectively\cite{Guggenheim1961, Karki2012}. Our transport measurements show that Pd-BST is also a SC with $T_{\rm c} \simeq$ 1 K, which indicates that PdTe$_2$ and PdTe still superconduct after heavy intercalation/substitution with Bi and Sb, albeit with a lower $T_{\rm c}$.

\subsection{dc transport properties of the sandwich Josephson junction.} 
To characterize how these TI nanowires are proximitized by the superconductivity in Pd-BST, we fabricated devices as shown in Fig. 2a in a false-color scanning electron microscope (SEM) image (for the transport characterization of the BST thin films used for device fabrication, see Supplementary Note 2 and Supplementary Fig. 3). These devices are SNS-type Josephson junctions (JJs) in which the TI nanowire is sandwiched by SCs. We can measure the supercurrent across the nanowire, but the 1D transport properties of the nanowire cannot be measured in these devices. Nevertheless, such a sandwich-JJ configuration allows us to quantitatively characterize the electronic transparency of the TI/SC interface. 
In the sample shown in Fig. 2a, the Pd pads to define the JJs had a width of 1 $\mu$m with the gap $L$ systematically changed from 100 to 150 nm; therefore, the BST nanowires in the JJs were 1-$\mu$m long and less than 150-nm wide (with 30-nm thickness). The data we present here were obtained on the second junction from the left (device 1) in Fig. 2a. For shorter $L$, we obtained shorted junctions, while for longer $L$ we obtained poor junctions with little or no supercurrent (see Supplementary Note 3 and Supplementary Fig. 4.).

We note that there is an important difference between a planar junction, that is usually realized in TI-based JJs, and the sandwich junction realized here; namely, Andreev bound states (ABSs) are formed directly between the two TI/SC interfaces in the latter, while in the former, the superconducting part that is relevant to ABSs is not the SC above the TI surface but is the proximitized portion of the surface covered by the SC\cite{Volkov1995, schuffelgen2019}. Even though planar junctions can have a large effective gap\cite{Ghatak2018, Bretheau2017}, they are effectively SS’NS’S junctions where S' can have a soft gap and/or be the source of additional in-gap states, whereas S’ can be minimized or even completely removed in sandwich junctions.
In the junctions realized here, the top and bottom surfaces of our BST nanowire remain well-defined, which suggests that they each form a SNS junction with the N region containing the topological surface states as treated by Fu and Kane\cite{Fu2008}. In such a line junction, one-dimensional Majorana bound states having a 4$\pi$-periodicity were predicted to show up, where time reversal symmetry is broken due to the phase difference across the junction.

The junction resistance $R$ of device 1 is plotted as a function of temperature $T$ in the inset of Fig. 2b. The sharp drop in $R$ at 1.16 K is due to the superconducting transition of Pd-BST. Zero resistance was observed only below $\sim$0.65 K, but a sharp rise in voltage $V$ in the current-voltage ($I$-$V$) characteristics allowed us to define the critical current $I_{\rm c}$ up to 0.8 K. At 20 mK, $I_{\rm c}$ showed a Fraunhofer-like pattern as a function of the perpendicular magnetic field (Supplementary Fig. 5a), giving evidence that the supercurrent is not due to a superconducting short-circuit (see Supplementary Note 4 for the data of other devices). The $I_{\rm c}$($T$) curve shown in Fig. 2b is convex, suggesting that this JJ is in the short-junction limit\cite{galaktionov2002}.
A theoretical fit\cite{galaktionov2002, schuffelgen2019} (red solid line) to the data yields the transparency $\mathcal T_{\rm crit} \simeq$ 0.7 with the transition temperature $T_{\rm c}$ = 1.15 K that is consistent with the $T_{\rm c}$ of Pd-BST.

When SNS-junctions have a high interface transparency, electrons can be Andreev-reflected multiple times at the SN interface without losing coherence. Such a process is called multipe Andreev reflection (MAR), which is discernible in the $I$-$V$ characteristics. As shown in Fig. 2c, we observed several peaks in the plots of $dI/dV$ vs $V_{\rm dc}$ at various $T$, which is characteristic of MAR (for the data up to a higher $V_{\rm dc}$, see Supplementary Note 5 and Supplementary Fig. 6); at 20 mK, if we assign the index $n$ from 2 to 9 as indicated in the figure, the plot of the voltage at the peak ($V_n$) vs $1/n$ lies on a straight line (Fig. 2c inset), which is the key signature of MAR. From the slope of this linear fit, we obtain the superconducting gap $\Delta_{\rm SC}$ = 184 $\mu$eV. The MAR feature is observed with the index $n$ up to 9 and at temperatures up to 1.0 K, which points to a strong superconducting proximity effect. Knowing $\Delta_{\rm SC}$, we can estimate the coherence length $\xi = \frac{\hbar v_{\rm F}}{\pi \Delta_{\rm SC}}$ = 390 nm in the proximitized BST, where the Fermi velocity $ v_{\rm F}$ = 3.69$\times$10$^{5}$ ms$^{-1}$ reported for BST\cite{he2015} is used; this $\xi$ confirms the short-junction nature of our JJ.

By fitting the linear portion of the $I$-$V$ characteristic at high $V_{\rm dc}$ above $2\Delta_{\rm SC}$ with a straight line (red dashed line in Fig. 2d), one can estimate the excess current $I_{\rm e}$ = 0.815 $\mu$A; also, from the slope of this linear fit, the normal-state resistance $R_{\rm N}$ = 266 $\Omega$ is obtained. Using $I_{\rm c}$ = 0.374 $\mu$A at 20 mK, we obtain the $eI_{\rm c}R_{\rm N}$ product of 99.5 $\mu$eV, which gives the $eI_{\rm c}R_{\rm N}/\Delta_{\rm SC}$ ratio of 0.54. This is among the largest ratio reported for a TI-based JJ\cite{Ghatak2018,schuffelgen2019}. Furthermore, based on the Octavio-Tinkham-Blonder-Klapwijk theory\cite{Octavio1983, Flensberg1988}, the other ratio $eI_{\rm e}R_{\rm N}/\Delta_{\rm SC}$ = 1.18 gives the interface transparency $\mathcal T_{\rm OTBK} \simeq$ 0.8. This $\mathcal T_{\rm OTBK}$ is in reasonable agreement with the transparency $\mathcal T_{\rm crit}$ obtained from $I_{\rm c}(T)$ mentioned above. Therefore, all the indicators [i.e. $I_{\rm c}(T)$ behavior, MAR feature, $eI_{\rm c}R_{\rm N}/\Delta_{\rm SC}$ ratio, and excess current] demonstrate that the TI nanowire in our device is experiencing a strong superconducting proximity effect thanks to a high interface transparency.

%
%
\subsection{Shapiro response.} 
When a phase-locked conventional JJ having a $2\pi$-periodic current-phase relation (CPR) is irradiated with an rf wave of frequency $f$, Shapiro steps\cite{shapiro1963} show up at quantized voltages $V_m$ equal to $\frac{mhf}{2e}$, where $h$ is the Planck constant and the step index $m$ is an integer. However, if topological Majorana bound states exist in a JJ, they contribute a $4\pi$-periodic CPR\cite{Fu2008} which leads to the disappearance of Shapiro steps with odd-integer $m$. The realistic situation where $2\pi$- and $4\pi$-periodic CPRs coexist has been theoretically examined\cite{dominguez2017}, and it was concluded that at lower frequency and at lower rf power, the $4\pi$-periodicity becomes more visible in terms of the missing Shapiro steps, even when the $4\pi$-periodic contribution is relatively small. 
For a JJ based on the 2D TI system HgTe, Bocquillon {\it et al.} reported\cite{bocquillon2017} missing steps with $m$ up to 9, but experiments based on 3D TIs found only the $m$ = 1 step to be missing\cite{wiedenmann2016, schuffelgen2019, Calvez2019, Ronde2020, Rosenbach2021}. 
As shown in Fig. 3, the $m$ = 1 step is clearly missing at rf frequencies up to 3.6 GHz in our device at 20 mK, while the first step becomes visible at 4.5 GHz. Wiedenmann {\it et al.} argued\cite{wiedenmann2016} that the $m$ = 1 step becomes missing only when $f$ is smaller than the characteristic frequency $f_{4\pi} \equiv 2eR_{\rm N}I_{4\pi}/h$, where $I_{4\pi}$ is the amplitude of the $4\pi$-periodic current. The crossover frequency of our device, around 4 GHz, suggests $I_{4\pi}\simeq$ 30 nA. This gives the ratio $I_{\rm 4\pi}/I_{\rm c} \simeq$ 0.08. 

The missing first Shapiro step is expected to be gradually recovered with increasing temperature, because thermal excitation causes quasiparticle-poisoning of the Majorana bound states and smears the $4\pi$-periodicity. As shown in Figs. 4a--4c for 3.6 GHz, this is indeed observed in our device, giving support to the topological origin of the effect\cite{wiedenmann2016, schuffelgen2019, Calvez2019, Ronde2020, Rosenbach2021}. 

It is prudent to mention that the $m$ = 2 Shapiro step is also (partially) suppressed at low frequencies in our 20-mK data (Figs. 3a and 3b). This is due to the hysteretic $I$-$V$ characteristic of an underdamped JJ\cite{schuffelgen2019}. To rule out this mechanism as the origin of the missing $m$ = 1 step, we show 2.95-GHz data taken at 800 mK in Fig. 4d; here, one can see that even at a high enough temperature where the hysteretic $I$-$V$ behavior is gone (see Supplementary Note 6 and Supplementary Fig. 7), the $m$ = 1 step is still missing. We note that, in addition to taking the data of $dV/dI$ vs $I_{\rm bias}$ with the ac technique, we have further measured the straightforward $I$-$V$ characteristics with a dc technique to completely rule out any hysteresis at 800 mK.
We also note that it was argued that Landau-Zener transitions\cite{dominguez2012} can mimic the behavior expected for the $4\pi$-periodic contribution when the JJ has a very high transparency close to 1\cite{Dartiailh2021}, but the transparency of our JJ is not that high. Hence, the Shapiro-step data of our device strongly suggest that a $4\pi$-periodic contribution to the CPR exists in our proximitized TI nanowire. This gives possible evidence for topological Majorana bound states\cite{Fu2008}, although one can never nail them down with the Shapiro-step data alone. The observation of missing first Shapiro step was reproduced in two more devices as shown in the Supplementary Note 7 and Supplementary Fig. 8.

\section*{Discussion}

In our STEM studies, by comparing different samples with the Pd pads aligned differently with respect to the crystallographic axes of BST, we found that the diffusion speed of Pd inside BST is slower along the $\langle110\rangle$ direction than the $\langle1\bar{1}0\rangle$ direction (see Supplementary Note 1 and Supplementary Fig. 1). The JJs reported here were aligned such that Pd diffuses along the slower $\langle110\rangle$ direction. 
We presume that a major part of the Pd diffusion and the structure conversion occurs already during the thermal deposition of Pd due to the heat coming from the crucible, but the exact understanding of the process requires dedicated studies.

The results presented here demonstrate a route to realize robust proximity-induced superconductivity in a TI nanowire. The induced superconductivity presents possible evidence for a topological nature in terms of the $4\pi$-periodicity caused by Majorana bound states which disperse along the nanowire\cite{Fu2008}. Note that these dispersive Majorana bound states cannot themselves be used for topological qubits to encode quantum information, for which localized MZMs are required. In this regard, our superconducting TI nanowire platform would be suitable for realizing a recent theoretical proposal\cite{Legg2021} to employ gating to create MZMs that are protected by a large subband gap to make them more robust against potential fluctuations due to disorder. The present platform further gives us an additional tuning knob, that is, the phase difference between the two SCs  sandwiching the TI nanowire; by making a SQUID loop, one can tune the phase difference with a small magnetic field. Since it was argued\cite{Fornieri2019, Ren2019} for a non-topological semiconductor platform that topological superconductivity can be engineered within a JJ by combining such a phase difference with an in-plane magnetic field, it would be interesting to see the role of phase difference in our TI-nanowire platform. Obviously, the construction of our system offers a lot of tuning knobs and its fabrication-friendly nature opens up an exciting prospect for exploring topological mesoscopic superconductivity.

\begin{methods}

\subsection{Sample preparations.} 
(Bi$_{1-x}$Sb$_x$)$_2$Te$_3$ films were grown with a molecular beam epitaxy (MBE) technique on sapphire (0001) or InP (111)A substrates by co-evaporating high-purity Bi, Sb, and Te from Knudsen cells in a ultra-high vacuum chamber.
The ${\rm Sb}/({\rm Bi} + {\rm Sb})$ flux ratio was 0.84, 0.90, and 0.67 for film 1 (used for device 1), film 2 (used for devices 2 and 3), and the STEM sample, respectively. 
After the electron-beam lithography using a Raith Pioneer II system, 20-nm-thick Pd layer was deposited onto the BST film by thermal evaporation from an alumina crucible, and a lift-off process was performed to leave the designed Pd pads/electrodes. Before the Pd deposition, the surface of the BST film was cleaned by Ar-plasma etching (10 W for 2 min). 

\subsection{STEM experiments.}
The focused ion beam (FIB) system FEI Helios Nanolab 400s was used for preparing the Pd/BST/InP multilayer lamella specimens. Before cutting with Ga ions of FIB, a layer of carbon ($\sim$160 nm) and a layer of Pt ($\sim$2 $ \mu$m) were deposited on the sample surface to protect the specimen from damage. Plasma cleaning was carried out afterwards to remove the surface contamination. An FEI Titan 80-200 ChemiSTEM microscope, equipped with a Super-X EDX spectrometer and STEM annular detectors, was operated at 200 kV to collect HAADF images and EDX results. The convergence angle of the electron-beam probe was set to 24.7 mrad and the collection angle was in the range of 70--200 mrad. The Dr. Probe software package was used for HAADF image simulation\cite{barthel2018}. VESTA software was used for drawing the crystal structures.

\subsection{Device fabrication and measurements.}
Since the Pd-electrode pattern was deposited onto a continuous BST film, for JJ devices we partially dry-etched the BST films with Ar plasma (50 W for 3 min) to electrically separate the electrodes (the dark-blue area in Fig. 2a is the etched part).
The JJ devices were measured in dry dilution refrigerators (Oxford Instruments Triton200/400) with the base temperature of $\sim$20 mK. Both dc and low-frequency ac ($\sim$14 Hz) lock-in techniques were employed in a pseudo-four-terminal configuration. 
For the Shapiro-step measurements, rf wave was irradiated to the JJs by using a coaxial cable placed $\sim$2 mm above the device, and $dV/dI$ was measured with a lock-in while sweeping the bias current $I_{\rm bias}$ in the presence of the rf excitation; the voltage on the JJ was calculated by numerically integrating the data of $dV/dI$ vs $I_{\rm bias}$.

\end{methods}

\begin{addendum}

\item[Acknowledgments:] 
We thank F. Yang for his help at the initial stage of this work and H. Legg for helpful discussions. This project has received funding from the European Research Council (ERC) under the European Union's Horizon 2020 research and innovation programme (grant agreement No 741121) and was also funded by the Deutsche Forschungsgemeinschaft (DFG, German Research Foundation) under CRC 1238 - 277146847 (Subprojects A04 and B01) as well as under Germany's Excellence Strategy - Cluster of Excellence Matter and Light for Quantum Computing (ML4Q) EXC 2004/1 - 390534769. G.L. acknowledges the support by the Research Foundation - Flanders (FWO, Belgium), file nr. 27531 and 52751.

\item[Author contributions:] 
Y.A. conceived the project. M.B. fabricated the devices. A.B., G.L., A.U. and A.T. grew TI thin films. X.-K.W., M.L. and J.M. performed the STEM analyses. M.B. and J.F. measured the devices. Y.A., M.B. and J.F. analyzed the transport data. Y.A., M.B. and X.-K.W. wrote the manuscript with inputs from all authors.

\item[Competing Interests:] The authors declare no competing interests.
 
\item[Correspondence:] Correspondence and requests for materials should be addressed to Y.A.

\item[Data availability:] The data that support the findings of this study are available from the corresponding
author upon reasonable request.

\item[Publication note:]
This version of the article has been accepted for publication, after peer review, but is not the Version of Record and does not reflect post-acceptance improvements, or any corrections. The Version of Record is available online at: 
http://dx.doi.org/10.1038/s43246-022-00242-6

\end{addendum}

\begin{figure}[h]
\centering
\includegraphics[width=\textwidth]{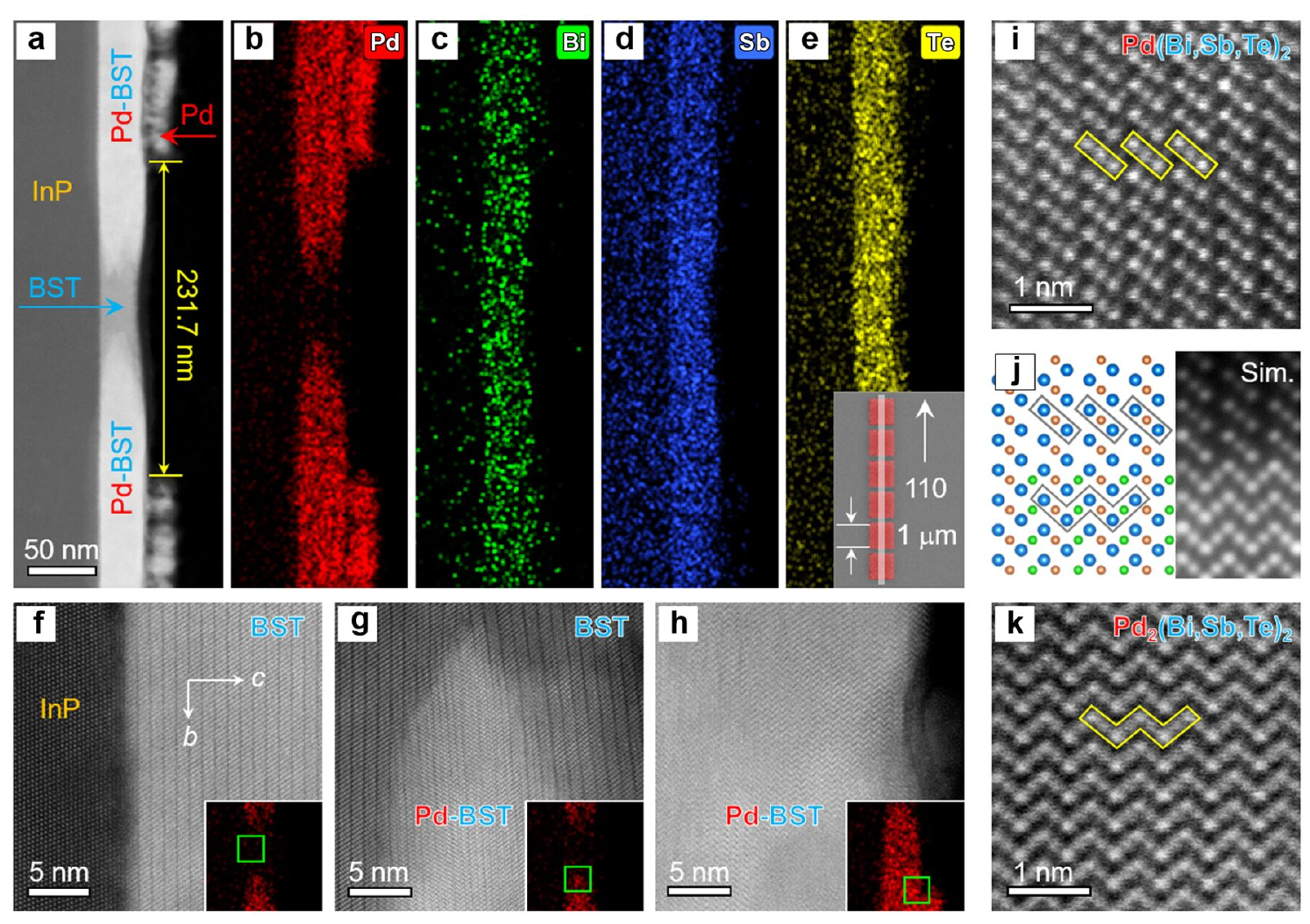}
\caption{\linespread{1.2}\selectfont{} \textbf{Cross-section of a BST nanowire sandwiched by Pd-BST.} \textbf{a-e,} HAADF-STEM image and corresponding EDX spectroscopy maps of Pd, Bi, Sb and Te, respectively, of a cross-section of the Pd-BST/BST/Pd-BST heterojunction on InP substrate. The inset in panel \textbf{e} shows a plan-view SEM image of the Pd-pad array; the light-gray stripe depicts how the lamella specimen shown in panel \textbf{a} was cut out. \textbf{f-h,} Atomic-resolution morphology of three locations in the specimen marked with green squares in the insets; the lattice image indicates the electron-beam incident direction of $\langle100\rangle$ or $\langle110\rangle$.
\textbf{i-k,} High-resolution HAADF image of PdTe$_2$-like (\textbf{i}) and PdTe-like (\textbf{k}) structure of Pd-BST, together with schematic structural model and simulated HAADF image (thickness $\sim$39 nm) to consider additional Pd atoms (green) intercalated into the PdTe$_2$-like structure (\textbf{j}). In panel \textbf{j}, the atomic sites coloured in blue, orange, and green may be occupied by Te/Bi/Sb, Pd/Sb, and Pd, respectively. Since the contrast of HAADF image is approximately proportional\cite{nellist2000} to square of the atomic number, $Z^2$, one can see from the atomic contrast in panel \textbf{i} that Bi atoms are found only on the Te sites of the PdTe$_2$ structure.
}
\label{fig:1}
\end{figure}

\begin{figure}[h]
\centering
\includegraphics[width=\textwidth]{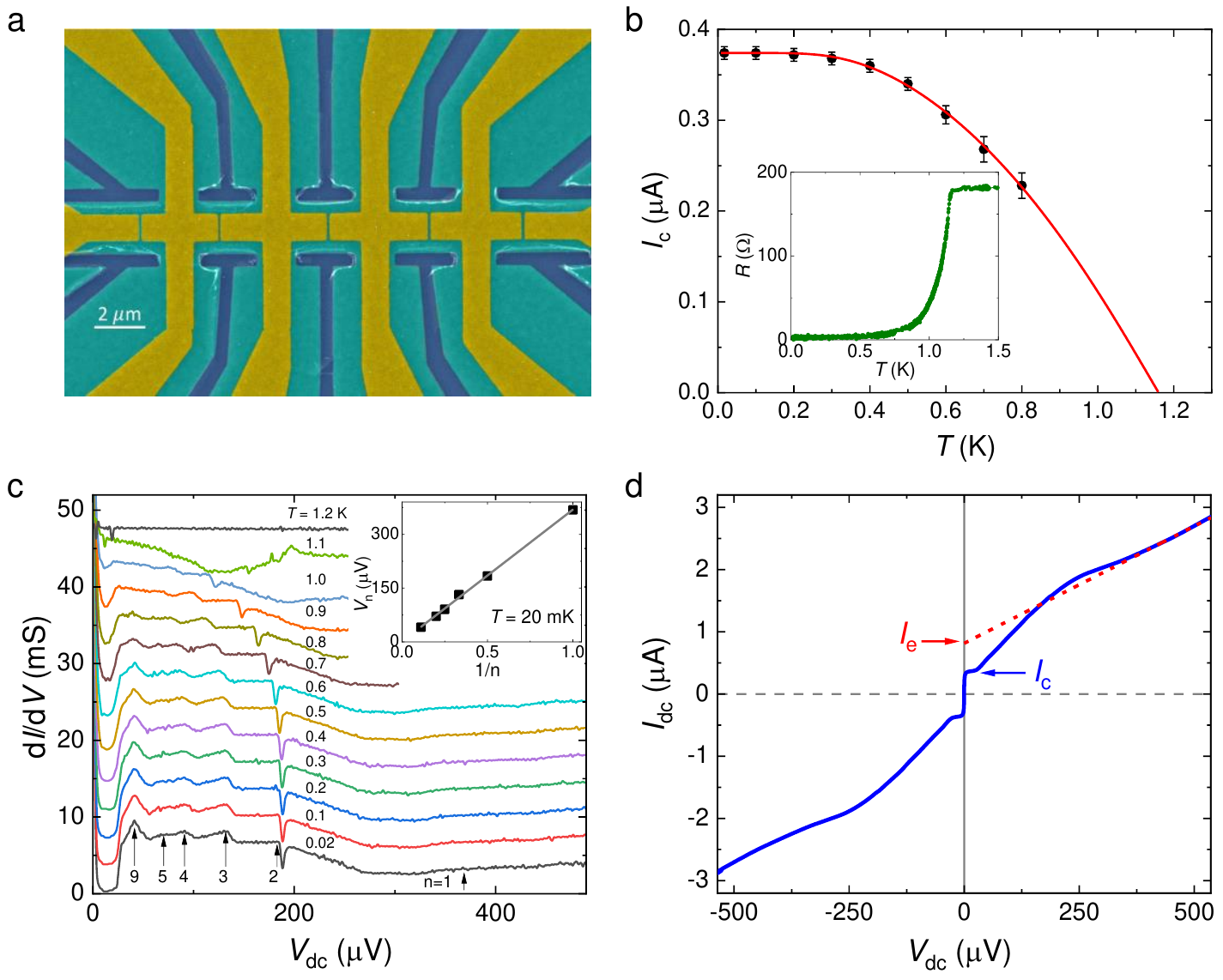}
\caption{\linespread{1.2}\selectfont{} \textbf{Properties of the Pd-BST/BST/Pd-BST sandwich-type Josephson junction.} \textbf{a,} False-color SEM image of the wafer including device 1 (2nd left junction); Pd electrodes (dark-yellow), BST film (turquoise), and sapphire substrate (dark-blue) are visible. The five junctions pictured here were fabricated with the Pd gap ranging from 100 to 150 nm (from left to right). \textbf{b,} $T$ dependence of the critical current $I_{\rm c}$ of device 1 with a theoretical fit (red solid line), where the error bars are the larger of (i) the amplitude of the ac excitation current ($\pm$5 nA) or (ii)  HWFM of the peak in $dV/dI$ vs $I_{\rm dc}$ at the breakdown of superconductivity; inset shows the $T$ dependence of the junction resistance measured with a quasi-four-terminal configuration. \textbf{c,} $dI/dV$ as a function of the dc bias voltage $V_{\rm dc}$ at various $T$; the curves are successively offset by 3.5 mS, except for the 20-mK data. Black arrows mark the MAR peaks with index $n$. Inset: The voltage corresponding to the MAR peaks, $V_n$, plotted vs $1/n$. \textbf{d,} $I$-$V$ characteristics at 20 mK in 0 T. A fit to the linear region at high bias (red dashed line) gives the excess current $I_{\rm e}$ and the normal-state resistance $R_{\rm N}$. Also, the $I_{\rm c}R_{\rm N}$ product can be obtained from these data.
}
\label{fig:2}
\end{figure}

\begin{figure}[h]
\centering
\includegraphics[width=\textwidth]{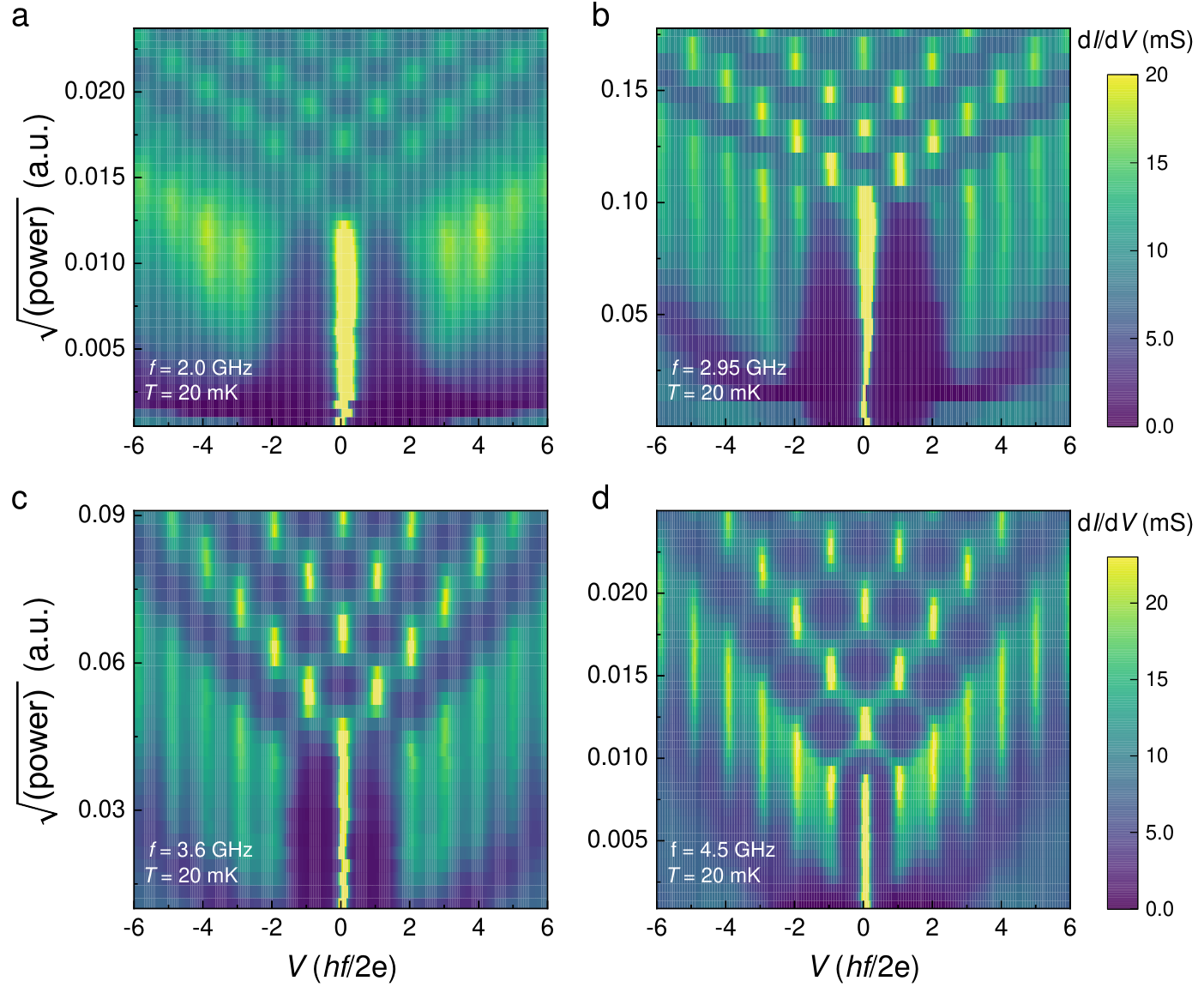}
\caption{\linespread{1.2}\selectfont{} \textbf{Shapiro response of device 1 at various rf frequencies.} \textbf{a-d,} Mapping of $dI/dV$ measured under the irradiation of rf waves at various frequencies (2.0, 2.95, 3.6, and 4.5 GHz) as functions of rf excitation and the dc voltage $V$ appearing on the JJ, which is normalized by $hf/(2e)$ to emphasize the Shapiro-step nature of the response. When there is a plateau (Shapiro step) at $V_m$ in the $V$ vs $I_{\rm bias}$ curve, the slope of the curve, $dV/dI$, becomes zero at the plateau, which means that $dI/dV$ diverges at $V_m$; therefore, the yellow vertical lines occurring at regularly-spaced $V$ is the signature of Shapiro steps.
}
\label{fig:3}
\end{figure}

\begin{figure}[h]
\centering
\includegraphics[width=\textwidth]{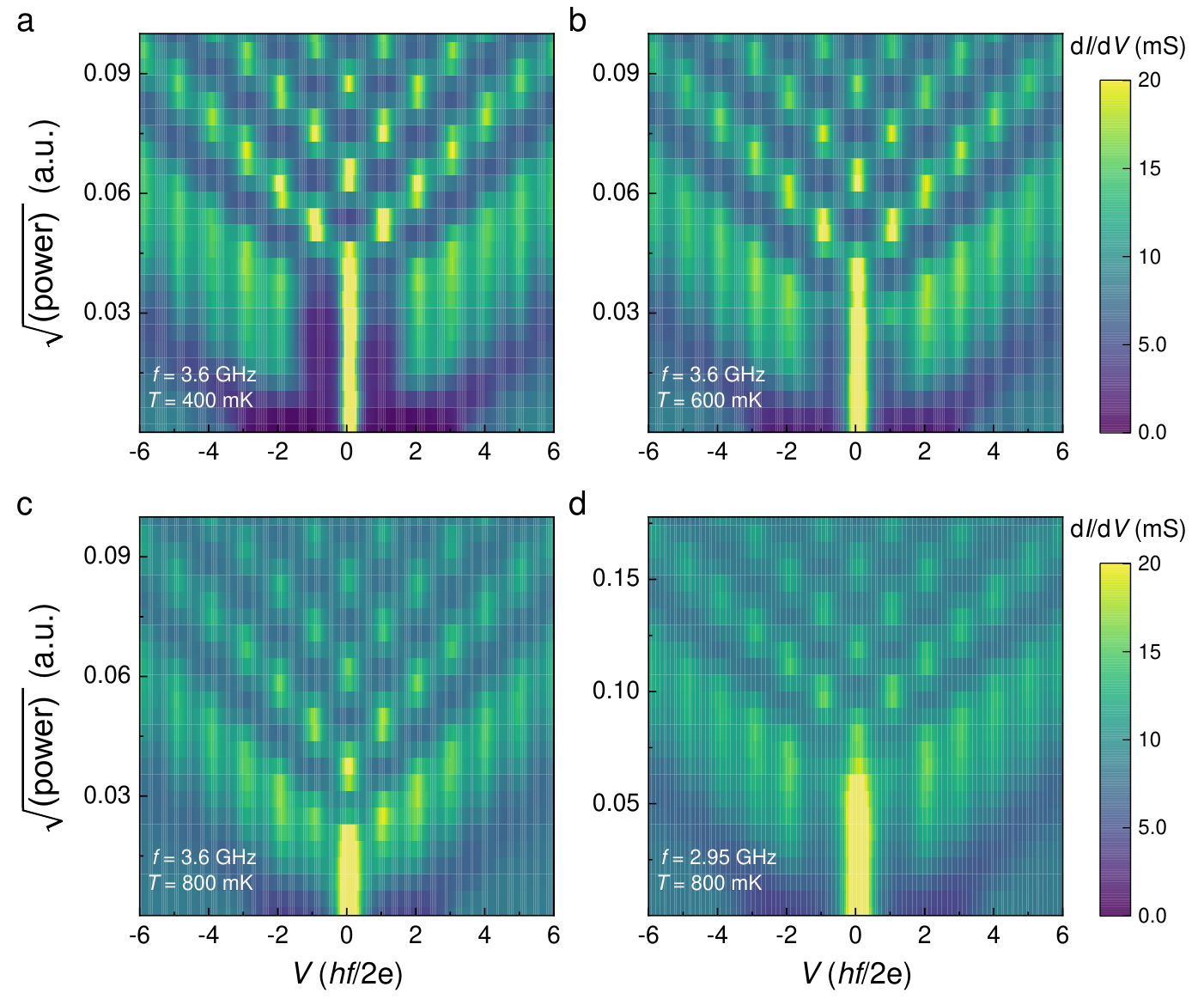}
\caption{\linespread{1.2}\selectfont{} \textbf{Shapiro response at higher temperatures.} \textbf{a-c,} Mapping of $dI/dV$  as functions of rf power and observed voltage for the rf excitation with 3.6 GHz (\textbf{a, b, c}) and 2.95 GHz (\textbf{d}) at higher temperatures. At 800 mK, the dc $I$-$V$ characteristics is completely free from hysteresis.
}
\label{fig:4}
\end{figure}

\clearpage

\renewcommand{\figurename}{Supplementary Figure}
\renewcommand{\theequation}{S\arabic{equation}}
\renewcommand{\thetable}{S\arabic{table}}

\renewcommand{\baselinestretch}{1.3}

\setcounter{figure}{0}
\setcounter{section}{0}

\begin{flushleft} 
\Large{\textbf{Supplementary Information}}
\end{flushleft}

\makeatletter
\let\saved@includegraphics\includegraphics
\renewenvironment*{figure}{\@float{figure}}{\end@float}
\makeatother

\large
\section*{Supplementary Note 1: Scanning Tunneling Electron Microscope (STEM) Analyses}
\normalsize

We performed STEM analyses on many Pd-BST/BST/Pd-BST heterojunction specimens with various Pd-pad distances as well as different Pd diffusion directions. Supplementary Figs. 1a--1e (top row) show the results of those specimens that were prepared such that the Pd diffusion across the Pd gap occurs along the $\langle110\rangle$ direction of the BST, while Supplementary Figs. 1f--1j (bottom row) show the results of other specimens prepared such that the diffusion occurs along the $\langle1\bar{1}0\rangle$ direction. Note that the hexagonal crystal symmetry of BST makes $\langle110\rangle$ and $\langle1\bar{1}0\rangle$ (that are orthogonal to each other) to be crystallographically inequivalent.

Representative cross-sectional morphologies of our $\langle110\rangle$-diffusion and $\langle1\bar{1}0\rangle$-diffusion specimens are shown in Supplementary Figs. 1a and 1f, respectively, as HAADF-STEM images. Importantly, one can see that a BST nanowire is formed between the two Pd-BST regions in Supplementary Fig. 1a, while the diffused Pd connects the two Pd-BST regions in Supplementary Fig. 1f. This means that diffusion constant of Pd atoms in BST is anisotropic, and the diffusion is faster along the $\langle1\bar{1}0\rangle$ direction than the $\langle110\rangle$ direction. This implies that a proper orientation engineering of the Pd array is crucial for the fabrication of BST nanowires. Our Josephson-junction (JJ) devices are made so that the Pd diffusion in the gap occurs along the slower $\langle110\rangle$ direction.

The EDX spectroscopy maps of Pd are shown for four specimens each for the $\langle110\rangle$-diffusion and $\langle1\bar{1}0\rangle$-diffusion cases in Supplementary Figs. 1b--1e and Supplementary Figs. 1g--1f, respectively. Here, the Pd-pad distance was systematically changed. The specimen shown in Supplementary Fig. 1c corresponds to the one reported in the main text.
In Supplementary Figs. 1b--1e, one can see that Pd atoms penetrate all the way down to the substrate in all specimens but always leave a Pd-free region (i.e. the BST nanowire) in-between the Pd-BST regions; in this series of specimens, the width of the nanowire increases from ca. 40 to 77 nm (Supplementary Figs. 1b--1e) as the Pd-pad distance increases from 191 to 239 nm. Remember that these STEM analyses were made about 3 months after the Pd deposition, so the nanowire width in a more ``fresh'' sample is expected to be larger.
For the $\langle1\bar{1}0\rangle$ direction, the Pd diffusion is obviously faster and difficult to control; for example, Supplementary Fig. 1j shows that even with the Pd-pad separation of $\sim$210 nm, Pd diffuses throughout the gap. 

To better understand the interface between BST and Pd-BST as well as the converted Pd-BST structure, Supplementary Figs. 2a--2c show additional, atomic-resolution HAADF-STEM images of the specimen shown in Supplementary Fig. 1a. The parallel triple-layer (TL) array near the ``epitaxial'' BST/Pd-BST interface reveals the dominance of the PdTe$_2$-like structure at the forefront of Pb-BST (Supplementary Fig. 2b). Note that the PdTe$_2$-like TL structure can be recognized by a dark contrast of the van-der-Waals gaps, while the PdTe-like structure gives a continuous lattice image.
In a region closer to the Pd pad (R2 in Supplementary Fig. 2a), the PdTe-like structure tends to dominates near the top surface (Supplementary Fig. 2c). This observation reproduces the result already described in the main text on a specimen with a wider Pd-pad gap. 
Similar structural behavior is also observed in a $\langle1\bar{1}0\rangle$-diffusion specimen (Supplementary Figs. 2d--2f). The structural models and simulated HAADF image are illustrated in the insets of Supplementary Figs. 2b, 2e, and 2f.

\large
\section*{Supplementary Note 2: Transport characterization of the (Bi$_{1-x}$Sb$_x$)$_2$Te$_3$ thin film}
\normalsize

We used two BST thin films grown on sapphire substrates for the device fabrications reported here. One (film 1) was used for device 1 and the other (film 2) was for devices 2 and 3, which were made to check for the reproducibility of the results (see later).
A part of the film was made into a Hall bar to characterize the transport properties. The measurements of the sheet resistance $R_{\rm S}$ vs temperature $T$, as well as the Hall resistance $R_{yx}$ vs magnetic field \textbf{B}, were performed in a Quantum Design Physical Properties Measurement System (PPMS) with the base temperature of 2--3 K. Supplementary Fig. 3a shows the $R_{\rm S}(T)$ and $R_{yx}$(\textbf{B}) behaviours of film 1, while Supplementary Fig. 3b shows those of film 2. The 2D carrier density of the films, $n_{\rm 2D}$, can be estimated from the slope of the $R_{yx}(B)$ data at \textbf{B} = 0 (which is equal to $R_{\rm H}/a$ with $R_{\rm H}$ the Hall coefficient and $a$ the film thickness) by using the formula $n_{\rm 2D}  = a (e R_{\rm H})^{-1}$, where $a$ is the film thickness. From the Hall data in the insets of Supplementary Figs. 3a and 3b, we obtain $n_{\rm 2D}$ of 2.05$\times$10$^{13}$ cm$^{-2}$ and 1.63$\times$10$^{13}$ cm$^{-2}$ for film 1 ($a$ = 30 nm) and film 2 ($a$ = 23 nm), respectively. 
The $R_{\rm S}(T)$ behavior and the $n_{\rm 2D}$ value of film 1 consistently show that this film is slightly bulk-conducting. On the other hand, in film 2, the $R_{\rm S}(T)$ behavior clearly show that this film is bulk-insulating, while the estimated $n_{\rm 2D}$ value is too large for a bulk-insulating film --- we often experience this kind of apparent inconsistency, which actually comes from the fact that the chemical potential is close to the Dirac point. Remember, when the chemical potential is swept across the Dirac point, the $R_{\rm H}$ value shows a zero-crossing\cite{Yang2015}, and a $R_{\rm H}$ value near this zero-crossing gives an apparent $n_{\rm 2D}$ value that is unphysically large. Therefore, one can conclude that our film 2 is bulk-insulating with the chemical potential relatively close to the Dirac point.


\large
\section*{Supplementary Note 3: Junctions with different gaps between Pd-pads}
\normalsize

We fabricated junction devices with a varying gap $L$ between the Pd-pads. As $L$ increases, we observed cross-overs from shorted junctions to proper SNS junctions and further to long junctions with little supercurrent. Typical superconducting transitions of the three types of junctions are exemplified in Supplementary Fig. 4a, and their $dV/dI$ vs $I_{\rm bias}$ curves are shown in Supplementary Fig. 4b. As one can see in Supplementary Fig. 4a, all three types of junctions have a superconducting transition at around 1 K, corresponding to the $T_c$ of the Pd-BST superconductor (i.e. PdTe$_2$ with Bi and Sb impurities), which we call $T_{\rm c, \mathrm{PdTe}_{2}}$. 

The resistance of a long junction remains finite until very low temperature, indicating that this is a weak SNS junction. Such a long junction has two critical currents as shown in Supplementary Fig. 4b; one is nearly zero and corresponds to the critical current of the JJ, $I_{\rm c, JJ}$, while the other is around 1.3 $\mu$A and corresponds to the critical current of the superconducting electrodes $I_{\rm c, \mathrm{PdTe}_{2}}$. No multiple Andreev reflection (MAR) feature is observed between $I_{\rm c, JJ}$ and $I_{\rm c, \mathrm{PdTe}_{2}}$ in a long junction, suggesting that junction is longer than the phase coherence length. In the case of a proper SNS junction, the resistance becomes zero at around 0.6 K;  $I_{\rm c, JJ}$ is a few hundred nA (which is much larger than in a long junction), and MAR features are clearly observed between $I_{\rm c, JJ}$ and $I_{\rm c,\mathrm{PdTe}_{2}}$. In a shorted junction, the resistance becomes zero immediately below $T_{\rm c, \mathrm{PdTe}_{2}}$ and there is only one critical current at $I_{\rm c, \mathrm{PdTe}_{2}}$; these observations indicate that PdTe$_{2}$ is present throughout the current path and no SNS junction is formed.

\large
\section*{Supplementary Note 4: Fraunhofer-like interference patterns}
\normalsize

To check for the reproducibility of our results, two more devices (2 and 3) were fabricated from another batch of the BST thin film (film 2), which was bulk-insulating as discussed above. The geometries of devices 2 and 3 were the same as that of device 1. 

When an out-of-plane magnetic field \textbf{B} is applied to a JJ, the critical current $I_{\rm c}$ presents a \textbf{B} dependence obeying the Fraunhofer interference pattern, $I_{\rm c}$(\textbf{B}) = $I_{\rm c0} | \frac{\sin (\pi \Phi/\Phi_0)}{\pi \Phi/\Phi_0}|$, where $\Phi$ is the total flux threading the junction and $\Phi_0 = \frac{h}{2e}$ is the magnetic flux quantum. Observation of the Fraunhofer-like pattern gives evidence that a Josephson junction is formed and the supercurrent is not due to some superconducting short-circuits. In the present case, this means that there is certainly a TI nanowire lying between the two Pd-BST parts.

Indeed, Fraunhofer-like patterns were observed in all three devices (Supplementary Fig. 5). Importantly, $I_{\rm c}$ drops to zero when $\Phi$ reaches $\Phi_0$ (first node), which is observed in all of them. However, since the Pd-BST portion of our devices was as wide as 1 $\mu$m, vortices are created in the Pd-BST in perpendicular magnetic fields and the flux jumps can spoil a clear Fraunhofer pattern. This flux-jump problem was particularly severe in device 1 even at low fields (Supplementary Fig. 5a), whereas cleaner Fraunhofer-like patterns were observed in devices 2 and 3 (Supplementary Figs. 5b and 5c). From the magnetic-field value of the first node, \textbf{B}$_1$, which was $\sim$3 mT in all three devices, we estimate the effective junction area $S_{\rm eff} =  \Phi_0$/\textbf{B}$_1 \simeq$ 0.7 $\mu$m$^{2}$. This area should correspond to $W(L_{\rm J}+2\lambda)$, where $W$, $L_{\rm J}$, and $\lambda$ are the junction width, junction length, and the London penetration depth, respectively. The estimated $S_{\rm eff}$ suggests $\lambda\ \simeq$ 300 nm.

\large
\section*{Supplementary Note 5: Absence of features above $2\Delta_{\rm SC}$}
\normalsize

An important point of the present work is that a sandwich junction is formed instead of a planar junction. To highlight the sandwich nature of our junctions, we show in Supplementary Fig. 6 the $I$-$V$ curve of device 1 at 20 mK plotted up to 0.62 mV. Here, $dI/dV$ calculated from the $I$-$V$ curve is also plotted. If the junction is a planar junction, one would expect\cite{kjaergaard2017transparent} additional features above 2$\Delta_{\rm SC}$. In our device, however, there is no clear feature above 2$\Delta_{\rm SC}$ besides the breakdown of the superconductivity in the PdTe$_{2}$ superconductor at  0.54 mV when $I_{\rm c,\mathrm{PdTe}_{2}}$ is reached. The linear fitting of the $I$-$V$ curve above 0.54 mV does not show an excess current, indicating that there is no superconductivity in the device above 0.54 mV.

\large
\section*{Supplementary Note 6: Missing Shapiro steps and hysteresis in the $I$-$V$ characteristics}
\normalsize

Hysteresis in the $I$-$V$ characteristics of a JJ can result in missing Shapiro steps. To rule out this possibility as the origin of our observation, we measured the differential resistance \textit{dV/dI} of device 1 for up and down current sweeps at various temperatures up to 0.8 K. As shown in Supplementary Fig. 7, a small hysteresis is observed up to 0.6 K, but at 0.8 K the hysteresis is gone. Nevertheless, as shown in Fig. 4d of the main text, the $m$ = 1 step is kept missing at 0.8 K. Hence, one can conclude that the missing first Shapiro step in device 1 is not due to hysteresis in the $I$-$V$ characteristics.

\large
\section*{Supplementary Note 7: Reproducibity of the missing first Shapiro step}
\normalsize

The missing of the first Shapiro step was reproduced in both devices 2 and 3. In device 2, the first Shapiro step was missing at 20 mK with the rf excitation at 2.6 GHz as shown in Supplementary Fig. 8a, where a weak asymmetry at lower power due to a hysteretic $I$-$V$ characteristics is visible; at 400 mK where the hysteresis is gone, the asymmetry is no longer observed while the first Shapiro step is still fully suppressed (Supplementary Fig. 8b). Device 3 did not show any hysteresis in the $I$-$V$ characteristics even at 20 mK and the first Shapiro step was missing at 20 and 400 mK with the rf excitation at 2.3 GHz (Supplementary Figs. 8c and 8d). The good reproducibility of the missing first Shapiro step gives confidence in the intrinsic nature of the $4\pi$-periodicity in our JJs involving BST nanowires.

\begin{figure}[h]

\centering
\includegraphics[width=\textwidth]{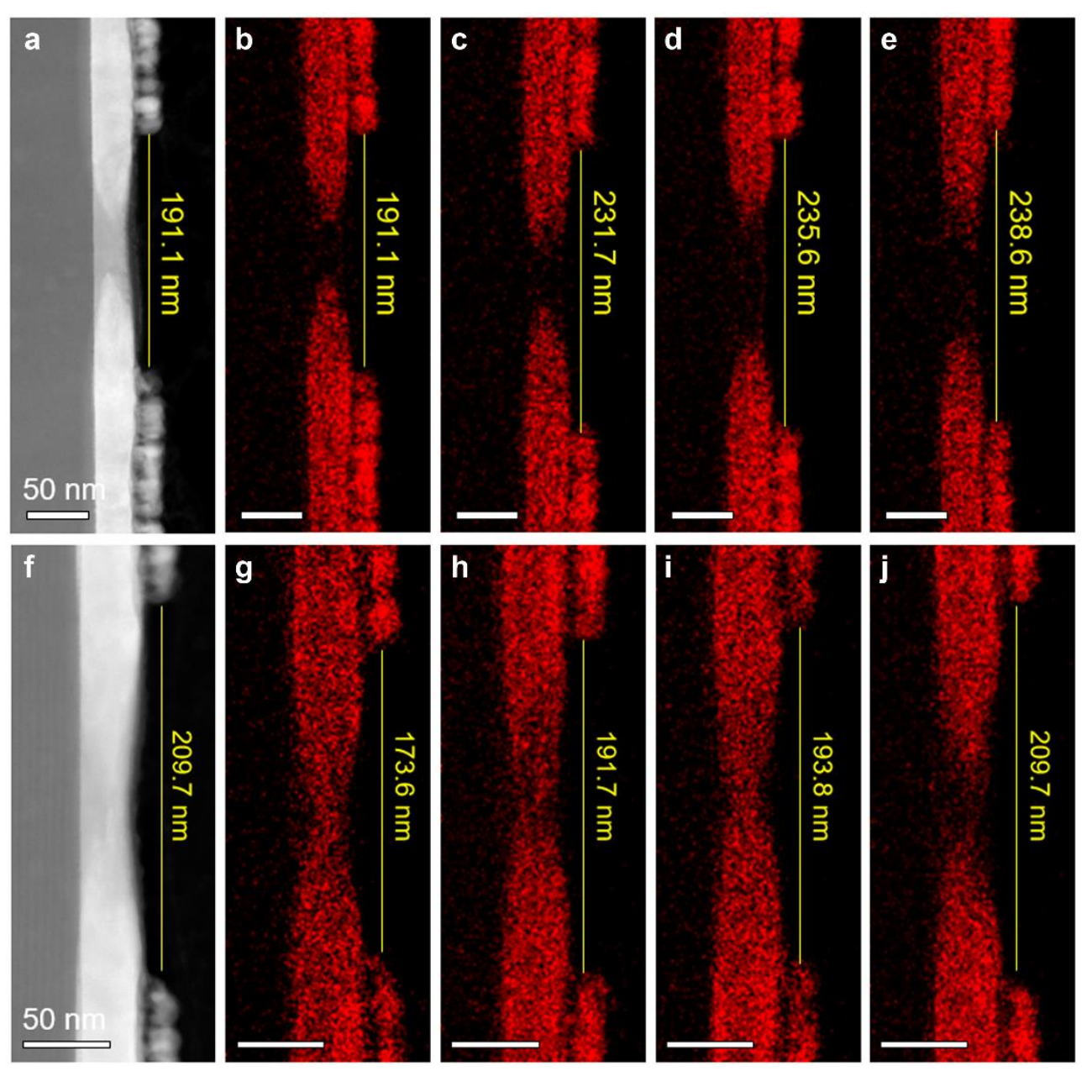}
\caption{\linespread{1.2}\selectfont{} \textbf{Visualization of Pd diffusion in the Pd-BST/BST/Pd-BST junctions.} 
\textbf{a,} HAADF-STEM image of a specimen, which was prepared so that Pd atoms diffuse in the $\langle110\rangle$ direction of BST. 
\textbf{b-e,} EDX spectroscopy maps of Pd in $\langle110\rangle$-diffusion specimens with various distances of the thermally-deoposted Pd pads. 
\textbf{f,} HAADF-STEM image of a $\langle1\bar{1}0\rangle$-diffusion specimen. 
\textbf{g-j,} EDX spectroscopy maps of Pd in $\langle1\bar{1}0\rangle$-diffusion specimens with various Pd-pad distances. 
}
\label{fig:S1}
\end{figure}

\begin{figure}[h]
\centering
\includegraphics[width=\textwidth]{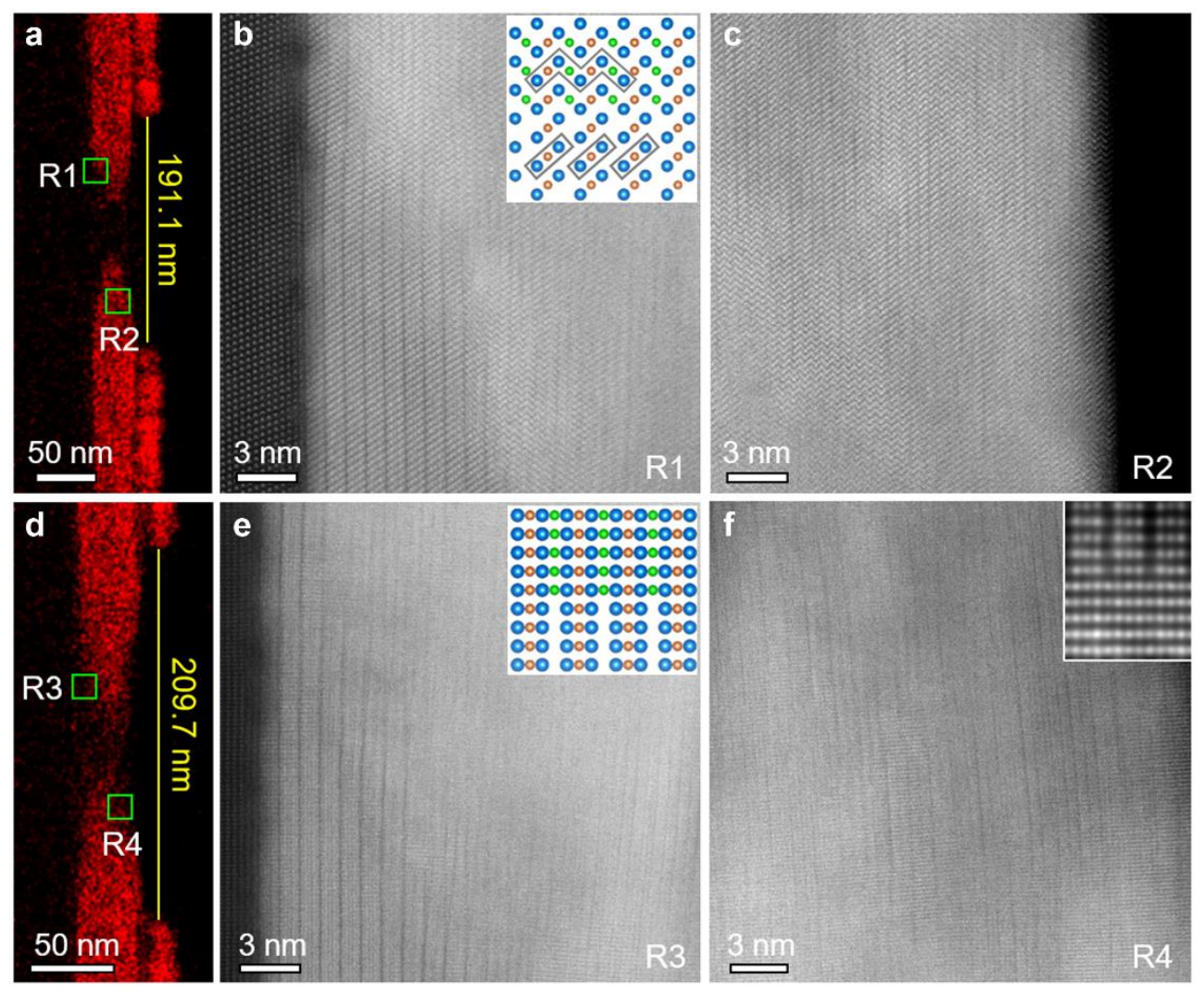}
\caption{\linespread{1.2}\selectfont{} \textbf{Structural analysis of the interface and the converted region.} \textbf{a-c,} EDX spectroscopy map of Pd in a $\langle110\rangle$-diffusion specimen (\textbf{a}, reproduction of Fig. S1b) and its atomic-resolution HAADF-STEM images (\textbf{b, c}) of the regions R1 and R2 indicated in panel \textbf{a} with green squares; the atomic arrangements in BST indicates electron-beam incident direction of $\langle100\rangle$ or $\langle110\rangle$.
\textbf{d-f,} EDX spectroscopy map of Pd in a $\langle1\bar{1}0\rangle$-diffusion specimen (\textbf{d}, reproduction of Fig. S1j) and its atomic-resolution HAADF-STEM images (\textbf{e, f}) of the regions R3 and R4 indicated in panel \textbf{d}; the atomic arrangements in BST indicates electron-beam incident direction of $\langle1\bar{1}0\rangle$.
The insets in panels \textbf{b} and \textbf{e} are schematic structural models of Pd-BST considering Pd intercalation into PdTe$_2$ and substitions by Bi and Sb; the atomic sites coloured in blue, orange, and green may be occupied by Te/Bi/Sb, Pd/Sb, and Pd, respectively. The inset in panel \textbf{f} is HAADF simulation of PdTe$_2$-like and PdTe-like phases of Pd-BST viewed along the $\langle1\bar{1}0\rangle$ direction (sample thickness $\sim$32 nm).
}
\label{fig:S2}
\end{figure}

\begin{figure}[h]
\centering
\includegraphics[width=\textwidth]{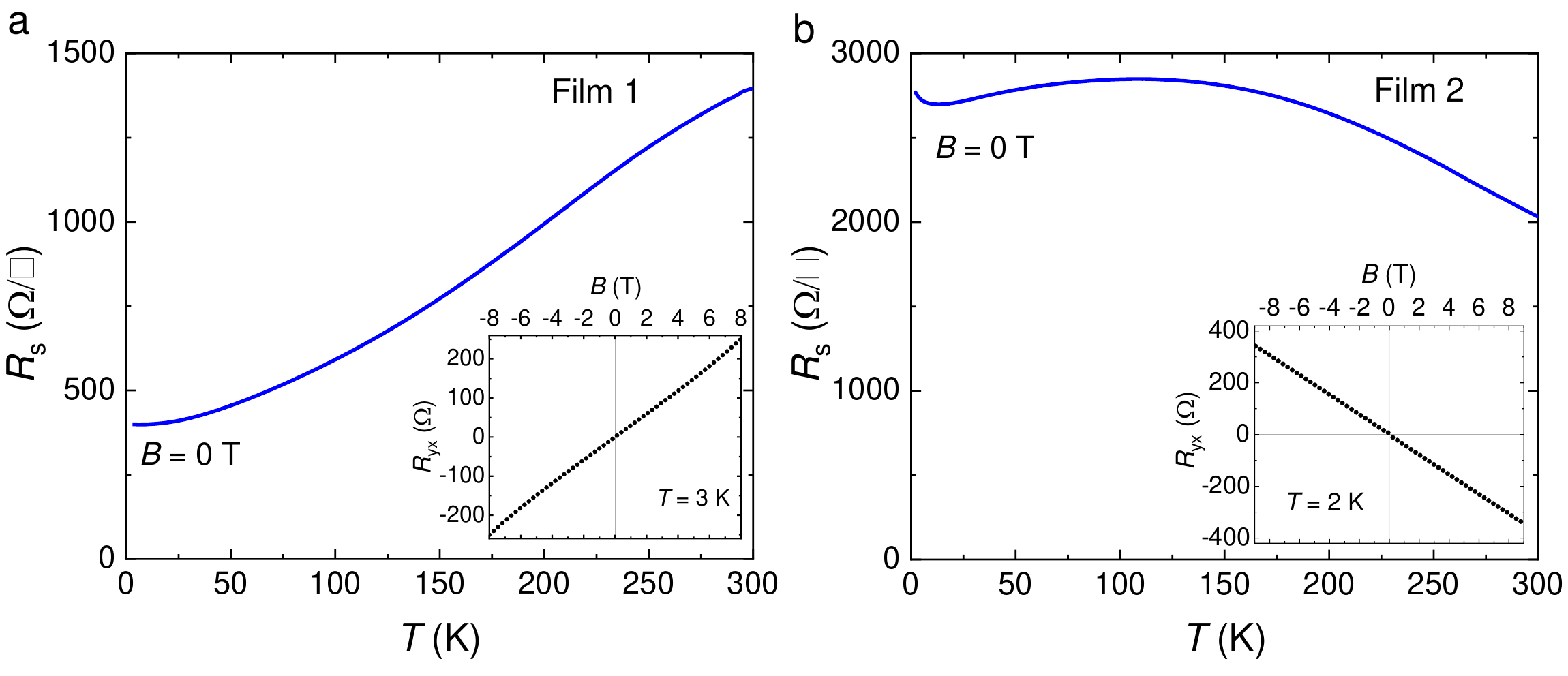}
\caption{\linespread{1.2}\selectfont{} \textbf{Transport properties of the BST films.} \textbf{a,b}, $R_{\rm S}(T)$ data for film 1 used for device 1 (\textbf{a}) and film 2 used for devices 2 and 3 (\textbf{b}). Insets show their $R_{yx}(B)$ data at the base temperature.
}
\label{fig:S3}
\end{figure}

\begin{figure}[h]
\centering
\includegraphics[width=0.7\textwidth]{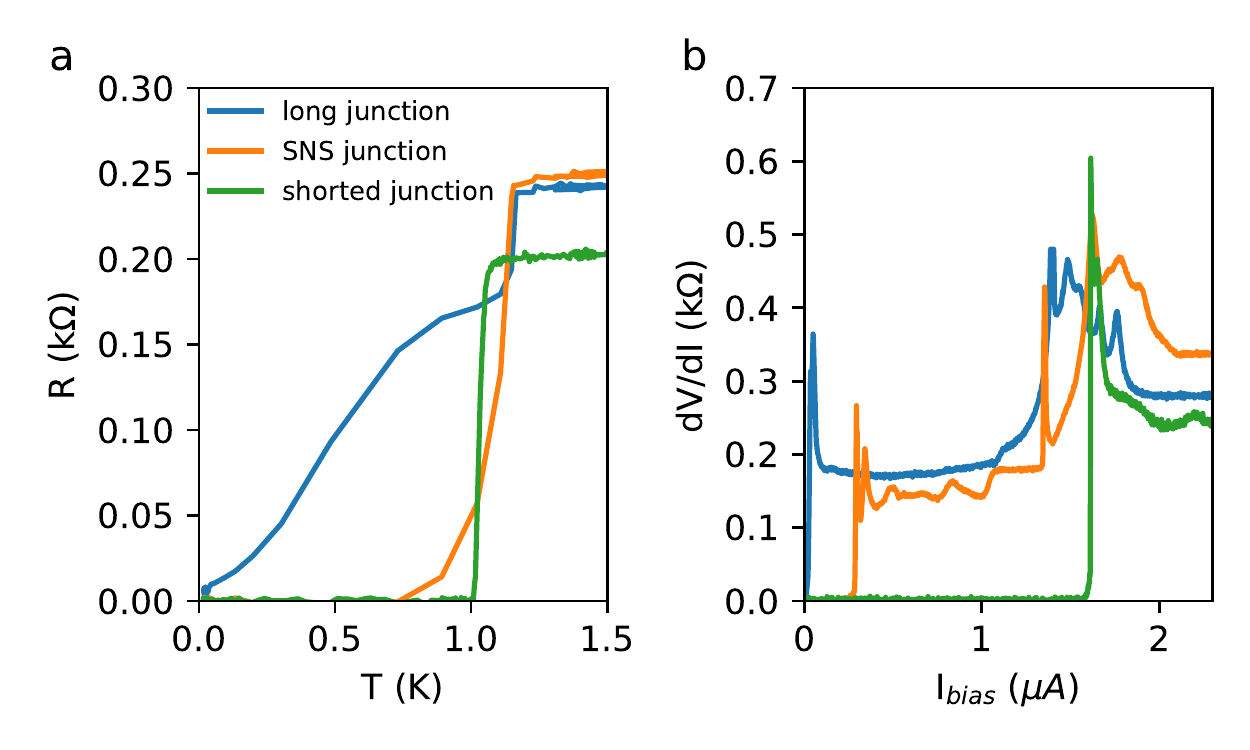}
\caption{\linespread{1.2}\selectfont{} \textbf{Three types of junctions.} \textbf{a,} Temperature dependences of the junction resistance for a long junction, a proper SNS junction, and a shorted junction. The data for a long junction and a SNS junction are from the same BST film, while the data for a shorted device is from another batch. \textbf{b,} Plots of $dV/dI$ of the three junctions shown in \textbf{a}, measured at 20 mK. 
}
\label{fig:S7}
\end{figure}

\begin{figure}[h]
\centering
\includegraphics[width=0.75\textwidth]{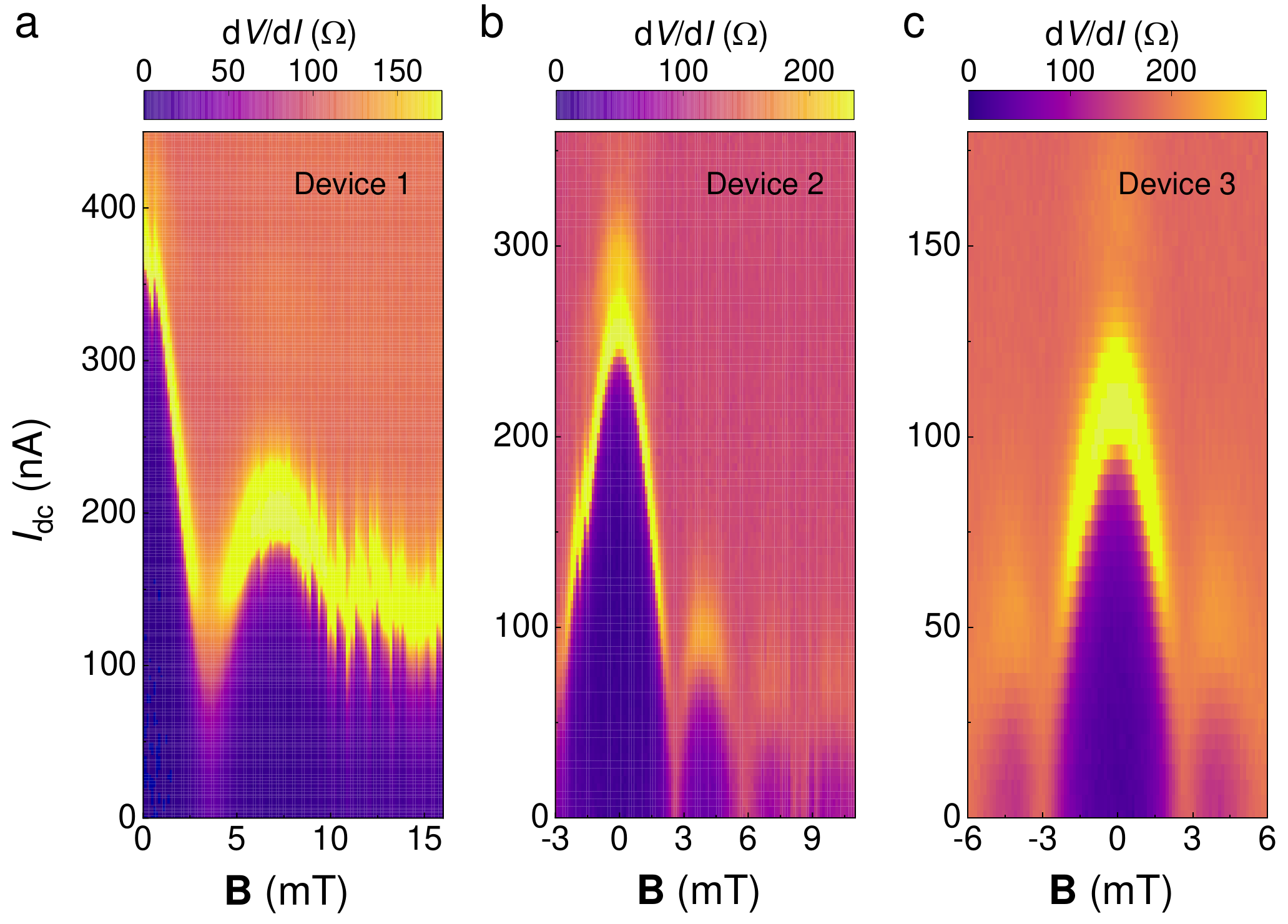}
\caption{\linespread{1.2}\selectfont{} \textbf{Fraunhofer-like patterns.} \textbf{a-c,} Mapping of the differential resistance $dV/dI$ as functions of the magnetic field $B$ and the dc current bias $I_{\rm dc}$ for devices 1, 2 and 3 measured at 20 mK. The discontinuous behaviour visible in panel $\textbf{a}$ is due to flux jumps. 
}
\label{fig:S4}
\end{figure}

\begin{figure}[h]
\centering
\includegraphics[width=0.7\textwidth]{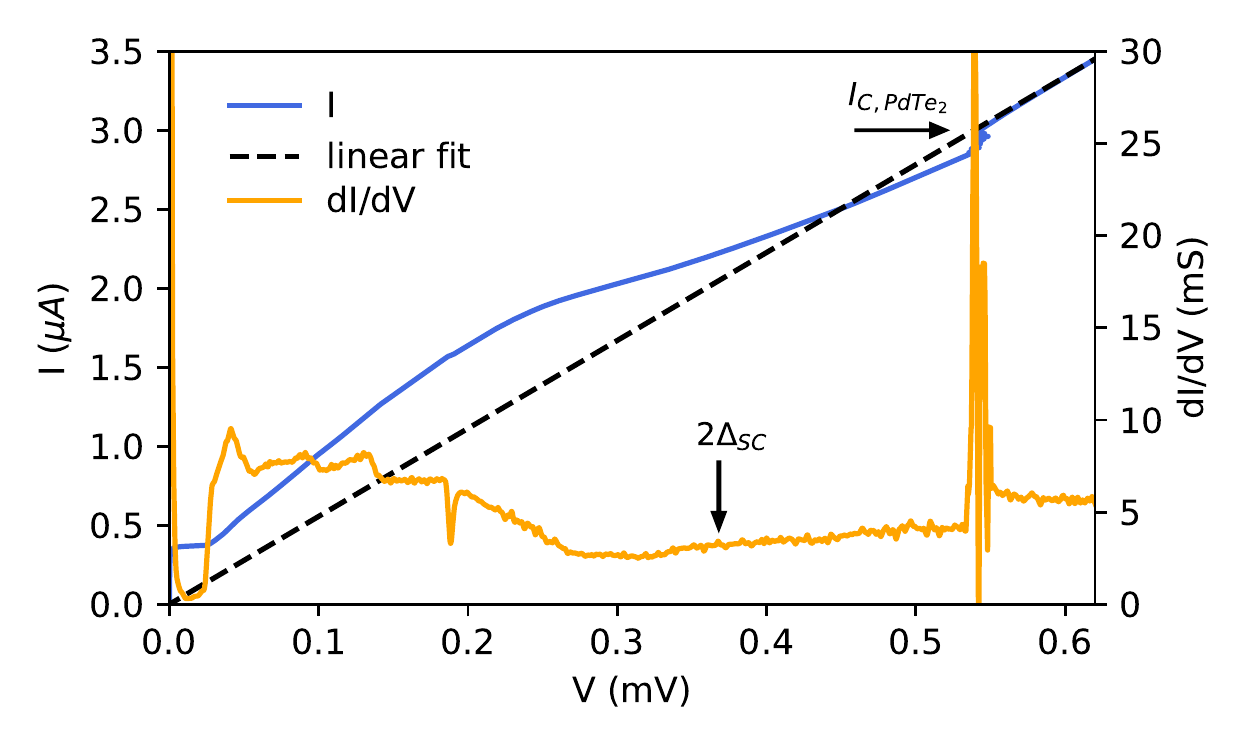}
\caption{\linespread{1.2}\selectfont{} \textbf{$I$-$V$ characteristic of device 1.} Plots of the $I$-$V$ curve (blue, left axis) and the $dI/dV$ vs $V$ curve (orange, right axis) of device 1 at 20 mK. The dashed line is a linear fitting of the $I$-$V$ curve above the critical current of the PdTe$_{2}$ electrodes. 
}
\label{fig:S8}
\end{figure}

\begin{figure}[h]
\centering
\includegraphics[width=0.75\textwidth]{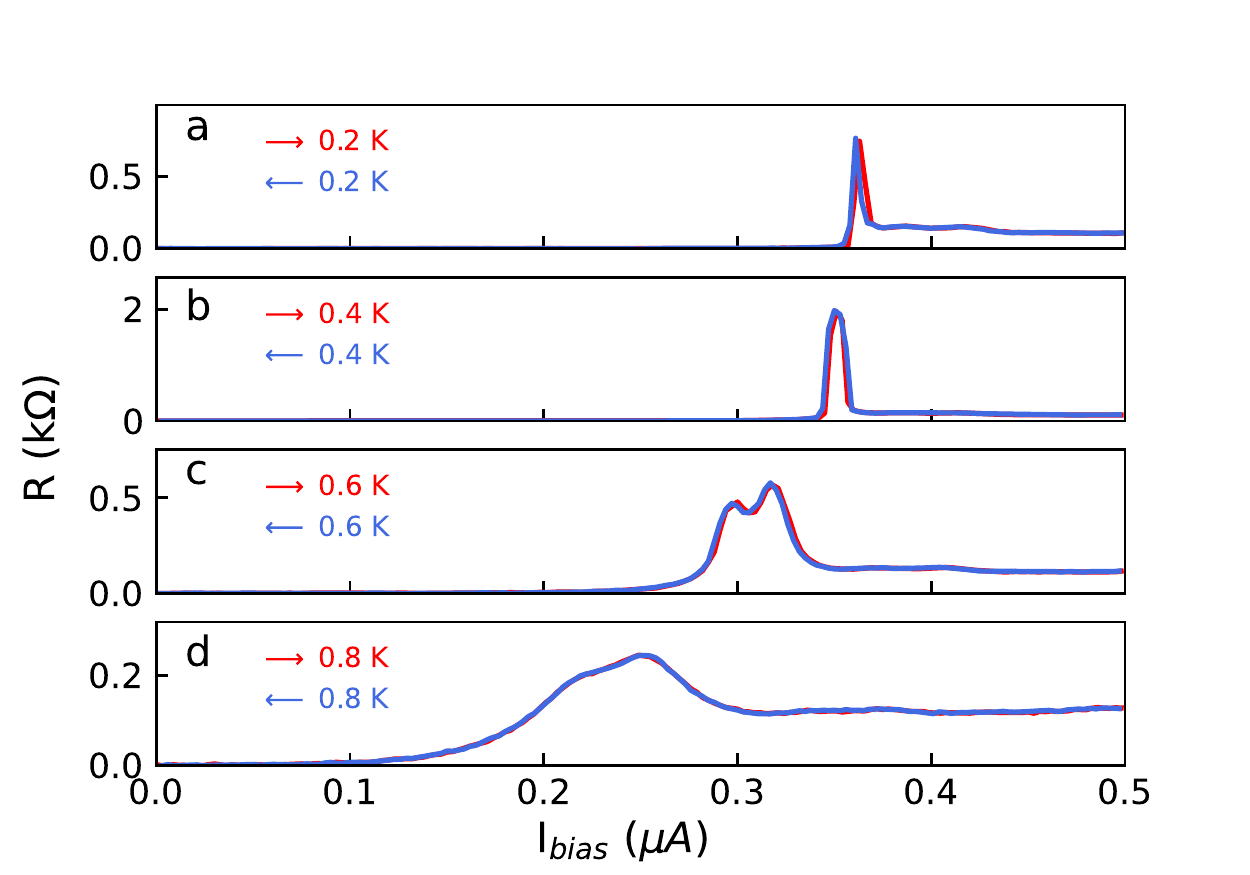}
\caption{\linespread{1.2}\selectfont{} \textbf{Hysteresis in the current response.} \textbf{a-d,} Plots of $dV/dI$ vs bias current $I_{\rm bias}$ at 0.2, 0.4, 0.6, and 0.8 K. Red curves are sweep-up and blue curves are sweep-down. All curves were measured without magnetic field nor rf excitation. 
}
\label{fig:S6}
\end{figure}

\begin{figure}[h]
\centering
\includegraphics[width=\textwidth]{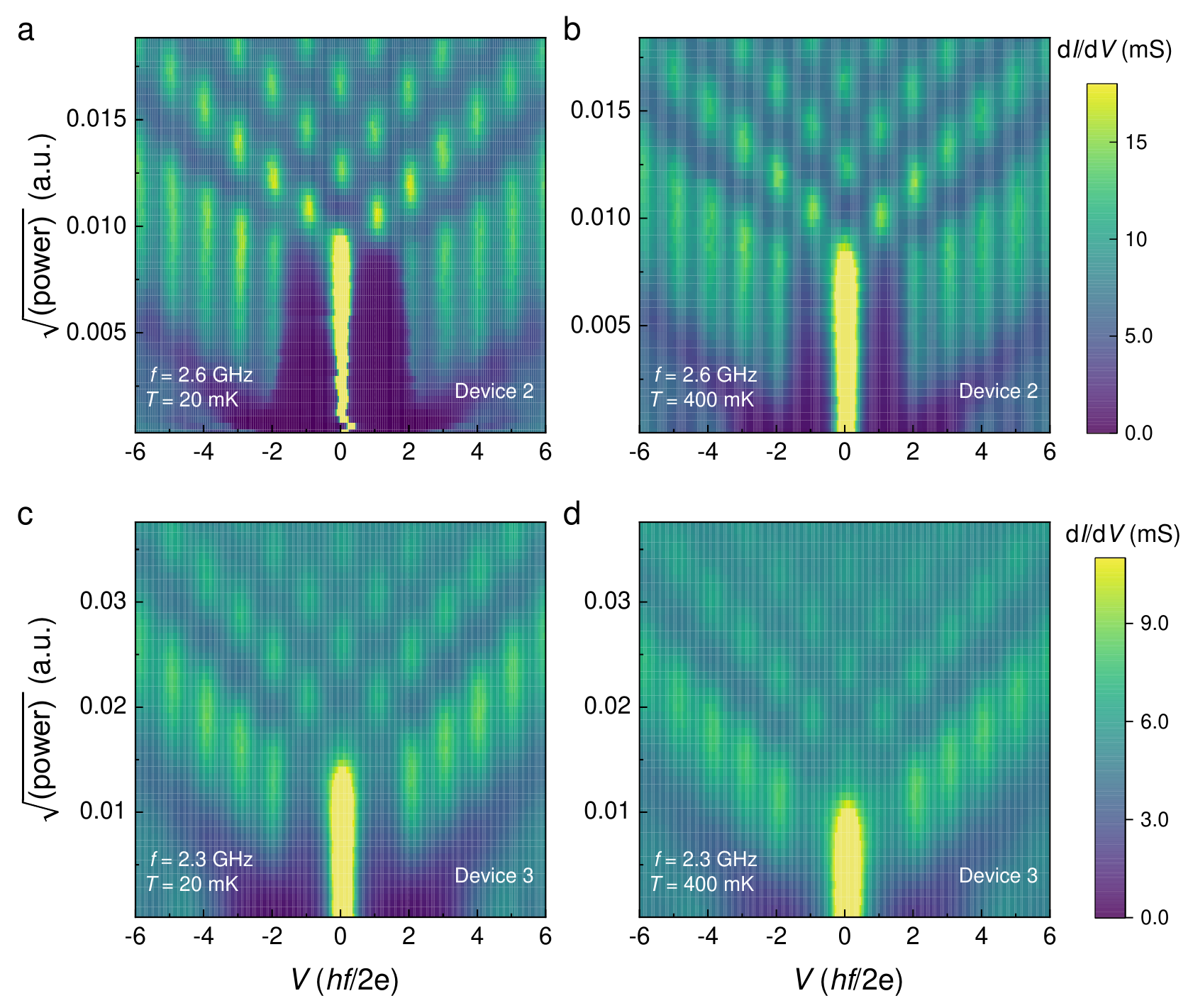}
\caption{\linespread{1.2}\selectfont{} \textbf{Shapiro response of devices 2 and 3.} \textbf{a,b,} Mapping of $dI/dV$ measured on device 2 at 20 and 400 mK, under the irradiation of rf waves at 2.6 GHz, as functions of rf excitation and the dc voltage $V$ appearing on the JJ, which is normalized by $hf/(2e)$ to emphasize the Shapiro-step nature of the response. \textbf{c,d,} Mapping of $dI/dV$ measured on device 3 at 20 and 400 mK with 2.3 GHz rf waves as functions of rf excitation and the dc voltage $V$.
}
\label{fig:S5}
\end{figure}

\vspace{-1mm}
\section*{Supplementary References}

\end{document}